\documentclass[,conference]{IEEEtran}
\pdfoutput=1
\usepackage[utf8]{inputenc}
\usepackage[usenames, dvipsnames]{color}
 \usepackage{booktabs, multicol, multirow}
\ifCLASSINFOpdf
  \usepackage[pdftex]{graphicx}
 \DeclareGraphicsExtensions{.pdf,.jpeg,.png}
\else
 \usepackage[dvips]{graphicx}
 \DeclareGraphicsExtensions{.eps}
\fi
\usepackage{amsmath}
\usepackage{algorithmic}
\usepackage{array}
\usepackage{tabu}
\ifCLASSOPTIONcompsoc
 \usepackage[caption=false,font=normalsize,labelfont=sf,textfont=sf]{subfig}
\else
\usepackage[caption=false,font=footnotesize]{subfig}
\fi
\graphicspath{{./figures/}}
\usepackage{flushend}
\usepackage{url}
\IEEEoverridecommandlockouts
\begin{document} 
\title{"420 Friendly": Revealing Marijuana Use via Craigslist Rental Ads$^1$}



\author{\IEEEauthorblockN{Anh Nguyen\IEEEauthorrefmark{1},
Long Nguyen\IEEEauthorrefmark{1},
Dong Nguyen\IEEEauthorrefmark{1}, 
Uyen Le\IEEEauthorrefmark{2}, and  
Tuan Tran\IEEEauthorrefmark{2} 
 }
\IEEEauthorblockA{\IEEEauthorrefmark{1}Saolasoft Inc., Centennial, CO 880211, USA\\
Email: {\{anguyen, lnguyen, dnguyen\}@saolasoft.com}}


\IEEEauthorblockA{\IEEEauthorrefmark{2}Sullivan University, Louisville, KY 40205,  USA\\
Email: {\{ttran, ule\}@sullivan.edu}}
\vspace{-0.3in}
\thanks{$^1$ The paper will be presented in 2017 Joint Workshop on Health Intelligence (W3PHIAI 2017)}

}

\maketitle

\begin{abstract}

Recent studies have shown that information mined from Craigslist can be used for informing public health policy or monitoring risk behavior. This paper presents a text-mining method for conducting public health surveillance of marijuana use concerns in the U.S. using online classified ads in Craigslist. We scraped more than 200 thousands of rental ads in the housing categories in Craigslist and devised text-mining methods for efficiently and accurately extract rental ads associated with concerns about the uses of marijuana in different states across the U.S. We linked the extracted ads to their geographic locations and computed summary statistics of the ads having marijuana use concerns. Our data is then compared with the State Marijuana Laws Map published by the U.S. government and marijuana related keywords search in Google to verify our collected data with respect to the demographics of marijuana use concerns. Our data not only indicates strong correlations between Craigslist ads, Google search and the State Marijuana Laws Map in states where marijuana uses are legal, but also reveals some hidden world of marijuana use concerns in other states where marijuana use is illegal. Our approach can be utilized as a marijuana surveillance tool for policy makers to develop public health policy and regulations. 

\end{abstract}
\IEEEpeerreviewmaketitle
\vspace{-0.2in}
\section{Introduction}
\label{sect:intro}

Marijuana was authorized for use in medical and recreational purposes in the states of Colorado and Washington in 2012 \cite{urlmmjstate}. The number is now increasing to twenty-five states and the District of Columbia which have laws legalizing marijuana in some form. Despite of some medical benefits,  over the past decade, researchers around the globe have documented the problems associated with marijuana use and driving for both youths and adults. For instance, the study of Fried et al. found that marijuana use reduces IQ by as much as eight points by age 38 among people who started using marijuana regularly before age 18 \cite{fried2002current}. Additionally, studies in \cite{dredze2012social,paul2015worldwide} also showed strong relationship between the use of marijuana with users' health harms including heart attack risks, lung irritants or cough, metal illness or birth weight decrease. Furthermore, data provided by the U.S. National Survey on Drug User and Health \cite{macleod2004psychological} indicated that youth with poor academic results were more than four times as likely to have used marijuana in the past year than youth with an average of higher grades. That said, marijuana poses considerable danger to the health and safety of the users themselves, their families, and our communities. Surveillance of actual marijuana demographics use and concerns would be very helpful for federal law and health policy makers to develop appropriate public health regulations.

Recently, mining data from online websites and social media for making public health policy, regulations or natural disaster responses has been conducted extensively. A cohort of users on online sharing platform like Facebook, Twitter, or Craigslist posting and discussing about a particular topic could suggest some level of interest in the topic of different communities. For example, the study of Wang et al. \cite{wang2014hurricane} using  Twitter corpus showed strong correlation of some hashtags and tweets posted by users living within the locations affected by the 2012 Hurricane Sandy with its movement. Additionally, utilizing Facebook platform, Schwartz et al. \cite{schwartz2014towards} developed models to predict the degree of depression of users from their Facebook data. Differently, Fries et al. developed a text-mining method to extract the demographics of Craigslist casual sex ads to inform public health policy \cite{fries2014mining}. Obviously, online social media and forums are great resources that we can utilize to understand the interests or concerns of different geographically located communities. However, accurately and efficiently extracting intelligence from unstructured texts and data in these platforms is challenging, and it has attracted significant research attention recently. 

Craigslist is a well-known website for classified advertisements across 700 cities and 70 countries \cite{url2016}. There are about 80 million classified ads posted each month in different sections devoted to jobs, housing, community, for sale, services, etc. The huge amount of data created by users in Craigslist everyday could suggest the interests or concerns of communities geographically located in different places thanks to the unique web structure of Craigslist, where ads are tagged with locations and of interest of only users living within the vicinity. In this paper, we examined the online Craigslist rental ads across 50 states of the U.S. to understand the concerns of marijuana use in different locations. We scraped rental ads in the housing categories in Craigslist and devised text-mining methods for efficiently and accurately extract rental ads associated with concerns about the use of marijuana in different states. We linked the extracted ads to their geographic locations and computed summary statistics of the ads having marijuana related concerns. Our data is then compared with the State Marijuana Laws Map published by the U.S. government in May 2016 and marijuana related keywords search in Google to verify our collected data with the demographics concerns of marijuana uses. Our data indicates strong correlations between Craigslist ads, Google search and the State Marijuana Laws Map in states where marijuana uses are legally allowed. More importantly, our data further shows some strong concern of marijuana uses in some states where the use of marijuana in any forms is illegal. The findings suggest that our approach can be utilized as a surveillance tool for monitoring marijuana uses in informing federal law or public health policy and regulation makers. 
\vspace{-0.1in}
\section{Data Collection and Methodology}
\subsection{Unique Ad Posting Structure in Craigslist}

Ads posted in Craigslist have special structure that allows us to link each collected ad to its geographic location. Particularly, Craigslist is organized in geographic location hierarchy based on countries, states, and cities. Each city has its own site consisting of different categories such as ``community'', ``housing'', ``personals'', ``for sale'', ``discussion forums'', etc., and subcategories such as ``apt/housing'', ``rooms/shared'', etc., in ``housing'' category. Our current research focuses on the ``housing" category, where the ads are most concerned about the uses of marijuana in any forms. Intuitively, this is because marijuana, if used, will directly affect to the involved parties (e.g., lessors and lessees) of the ads. 

In our research we also notice that besides the ``housing" category, listings with marijuana-related terms can be found in other categories as well. For example, one can find a marijuana related terms in a job ad for recruiting a contract labor or a startup company is looking for a ``420 friendly" blogger. Most of those listings suggest the pro-marijuana sentiment. Due to the page limit, the deep search expanding to those categories will be considered in our future work.

\begin{figure}[t]
\begin{center}
\includegraphics[width=2.7in]{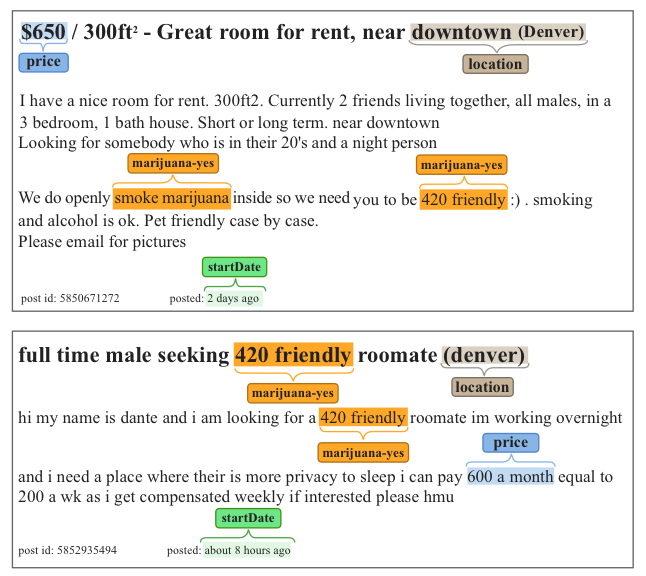}
\caption{\em Examples of Annotated Ads in Craigslist with Information of Marijuana Use Concern, Locations, etc.}
\label{fig:  cl_marijuana_refining}
\end{center}
\vspace{-0.3in}
\end{figure}
\vspace{-0.1in}
\subsection{History of Term ``420"}
Among marijuana smokers, the term ``420" (pronounced ``four-twenty") is familiar. Simply, it just means marijuana. The term has a long history back in 1971. Some high school students in San Rafael, California circulated a coded message started as ``420 Louis", meaning ``at 4:20 they'd meet by the Louis Pasteur statue outside the high school and get high." Years later, ``420" became California's police code indicating smoking pot. It then also has a meaning that 4:20 is the time to smoke pot and 4/20 is the international pot-smoking day \cite{420_url}. Today, the term ``420'' is used very often in marijuana-related circumstances, especially rental ads on Craigslist such as ``420 is ok", ``420 is allowed", or ``we are 420 friendly". 
\vspace{-0.1in}
\subsection{Data Reprocessing}

Extracting useful information from Craigslist ads is very challenging due to the unstructured ads. Fig. \ref{fig: cl_marijuana_refining} shows an example of free-format annotated ads in Craigslist. Despite of unstructured texts and hidden information, but if processing appropriately, we can extract a lot of useful information suggesting the interests of parties involved. For example, as annotated in Fig. \ref{fig: cl_marijuana_refining}, by taking into account the context and semantic of the posted corpora in housing categories we can extract significant useful information suggesting the interests or concerns of the involved parties. For example, we can extract not only price of the rent (i.e., \$650), but also its location (i.e., Denver), marijuana use concern (.i.e., 420 friendly - yes). Two main steps in our data preprocessing:
\begin{itemize}
\item {\it Duplicate and Outlier Removal}: We also notice that there are many duplicate ads posted in Craigslist despite of no-duplicate policy of Craigslist. Our first step in data preprocessing is to remove the duplicate ads. Those listings with exactly the same titles including price, location, number of bedrooms or those with different titles but the same content will be considered to be duplicate. Our data preprocessing step also filtered out all outlier ads with unrealistic prices.
 
\item {\it Extracting Marijuana Corpora}: Next, we used text-mining and semantic analysis algorithm to extract ads with having concerns about the use marijuana. The accuracy of corpora extraction in this step is very important as it determines our data analysis outcome. Interestingly, in our text processing we found out that ``420'' corpora such as ``420 friendly" or ``420 ok" are often used in the ads than the corpus ``marijuana'' itself. However, due to the unstructured texts, we need to filter out ``420' unrelated terms. For example, these terms ``\#420", ``APT 420", ``420 bucks per month", ``my number is xxx-420-xxxx'', etc. appear very often in housing ads but they should not be considered as marijuana corpora. Our developed text-mining and semantic algorithms have accurately filter out all of these false positive records.
\end{itemize}
\begin{figure*}[t!].
\begin{center}
\includegraphics[height = 1.15in]{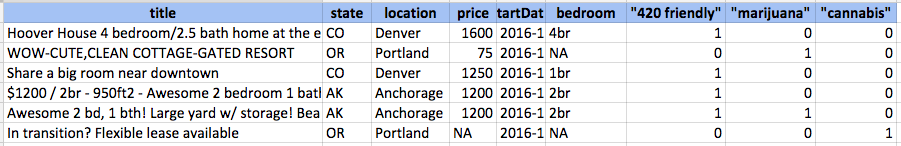}
\caption{\em A Sample of Dataset (``NA'' denotes not available data, ``0", ``1", ... are number of counts of the terms in the ad)}
\label{fig: dataset_sample}
\end{center}
\vspace{-0.2in}
\end{figure*}

\vspace{-0.12in}
\subsection{The Dataset}
Figure \ref{fig: dataset_sample} shows sample of our collected dataset across 50 states in the U.S. After removing the duplicate, outlier and unrelevant ads, we constructed a dataset of 200,000 rows and 9 columns where each row represents an ad and each column represents different information extracted from the ad including title of the ad, time listed, state, city, marijuana corpora, rent price, etc. In the dataset, we counted the number of occurrences of each marijuana corpus in different columns. We grouped all marijuana related corpora with ``420" keyword such as ``420 is ok", ``420 fine", ``is 420 allowed" to the field ``420 friendly".

\vspace{-0.05in}
\section{Results and Discussion}
\label{sect:discussion} 

Our extensive analysis based on the collected data provides some insights into marijuana use and concerns across different states in the U.S. Our data indicated significant differences of marijuana use concerns depending on the legal regulations associated with states and geographic locations. The main findings are summarized in the following subsections.

\subsection{Marijuana Corpora}

\begin{table}[t]
\begin{tabular}{ |p{3cm}||p{0.6cm}|p{0.6cm}|p{0.6cm}|p{0.6cm}|p{0.6cm}| }
 \hline
 \multicolumn{6}{|c|}{Marijuana Corpora Summary} \\
 \hline
 Category&$T1$&$T2$&$T3$&$T4$&$T5$\\
 \hline
rooms \& shares &1012&387&26&38&149\\
room/share wanted&418&46&35&55&19\\
apts/housing for rent&148&340&128&41&91\\
sublets \& temporary&108 &42&11&17&39\\
 \hline
 TOTAL &1686 &591&200&151&298\\
 \hline
\end{tabular}
\caption{\em Summary of Marijuana-Related Corpora in Subcategories of "housing" in the state of Colorado: $T1$ = ``420 friendly'',   $T2$ = ``marijuana'', $T3$ = ``mmj'', and $T4$ = ``cannabis'', $T5$ = ``pot"}
\label{tab: number_marijuana_listings}
\vspace{-0.2in}
\end{table}

In our data analysis, we found out that ``marijuana'' corpus are not the most often used terms to describe marijuana-related concerns (e.g., allowed or not allowed use of marijuana). This is a surprising result at first as based on the existing studies we expected ``marijuana'' related keywords should be the most often used terms. By further carefully examining all marijuana related terms, both formal and slang ones 
in the ``housing" category, surprisingly, we found that ``420" corpora (e.g., ``420 friendly'') are most often used in the rental ads to indicate marijuana use concerns, instead of ``marijuana" related keywords. More specifically, we found out that the ``420'' corpora used the most in two sub-categories ``room \& shares" and "room/share wanted", appear more than ten times as the ``marijuana'' related keywords as shown in Table \ref{tab: number_marijuana_listings} and about four times more than total of all other marijuana related keywords used. This finding is very interesting as it suggests a new term to indicate ``marijuana'' corpus when conducting text mining. 

Our result is a great complement to the existing studies on marijuana related aspects in online platforms such as \cite{cavazos2015twitter,daniulaityte2015time}. Those studies skip very important ``420" marijuana related keywords when investigating marijuana related content on Twitter.
\subsection{The Division of Rental World in Marijuana Related Ads}
Our study further revealed the difference of keyword usages in describing marijuana concerns between organization and private lessors. Based on our collected data, we compared the number of rental ads with and without marijuana-related terms from  sub-categories ``office \& commercial" in which listings are from rental companies. In this analysis, we only focused on four marijuana-legalizing states including Colorado, Washington, Oregon and Alaska. The data indicates that rental companies tend to use formal keywords such as ``marijuana" or ``cannabis" while the private parties used ``420'' related keywords more often as indicated in Table \ref{tab: number_marijuana_listings}. Table \ref{tab: with_without_420} shows the total number of posts containing term ``marijuana", ``cannabis", and ``420" related term in four states within one month.

\begin{table}[]
\centering
\begin{tabular}{|l|l|l|l|}
\hline
\multicolumn{1}{|c|}{\multirow{2}{*}{States}} & \multicolumn{3}{c|}{office \& commercial sub-category} \\ \cline{2-4} 
\multicolumn{1}{|c|}{}                        & ``marijuana"  & ``cannabis" & ``420 friendly" \\ \hline
Colorado                                      & 223           & 68          & 14              \\
Washington                                    & 181           & 29          & 13              \\
Oregon                                        & 497           & 303         & 7               \\
Alaska                                        & 23            & 25          & 2               \\ \hline
\end{tabular}
\caption{\em Summary of Marijuana-Related Corpora  in Subcategory ``office \& commercial"}
\label{tab: with_without_420}
\vspace{-0.15in}
\end{table}

\begin{figure}[t]
\begin{center}
\includegraphics[height = 1.3in, width = 2.79in]{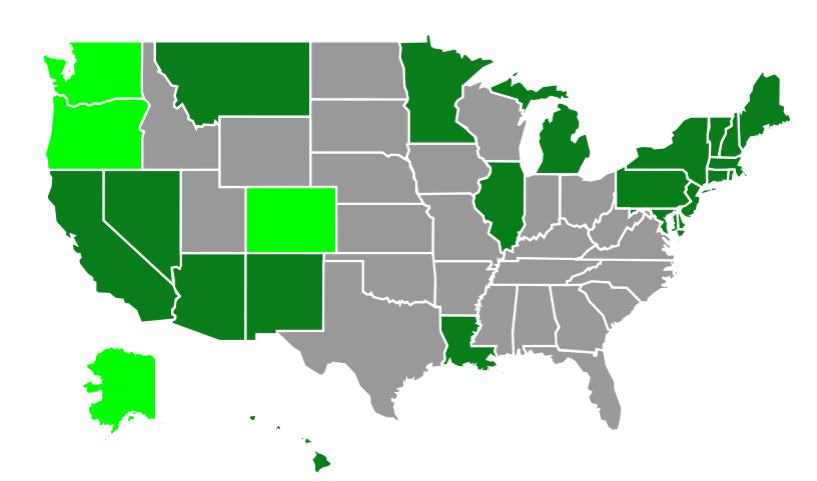}\\ (a)\\
\includegraphics[height = 1.3in, width = 2.79in]{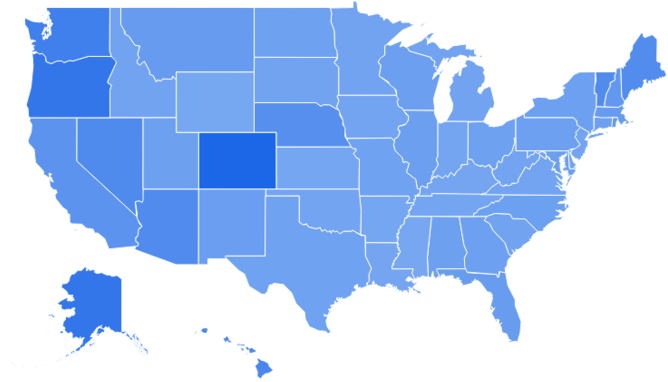} \\(b) \\
\caption{\em Comparison of (a) state marijuana laws map and (b) geographical distribution of marijuana related corpora ads in Craigslist (number of listings per population)}
\label{fig: states_with_4201}
\end{center}
\vspace{-0.3in}
\end{figure}
\subsection{Comparing to the Marijuana State Laws Map and Google Marijuana Keyword Search}
\begin{figure}[t]
\begin{center}
\includegraphics[height = 1.0in, width = 2.79in]{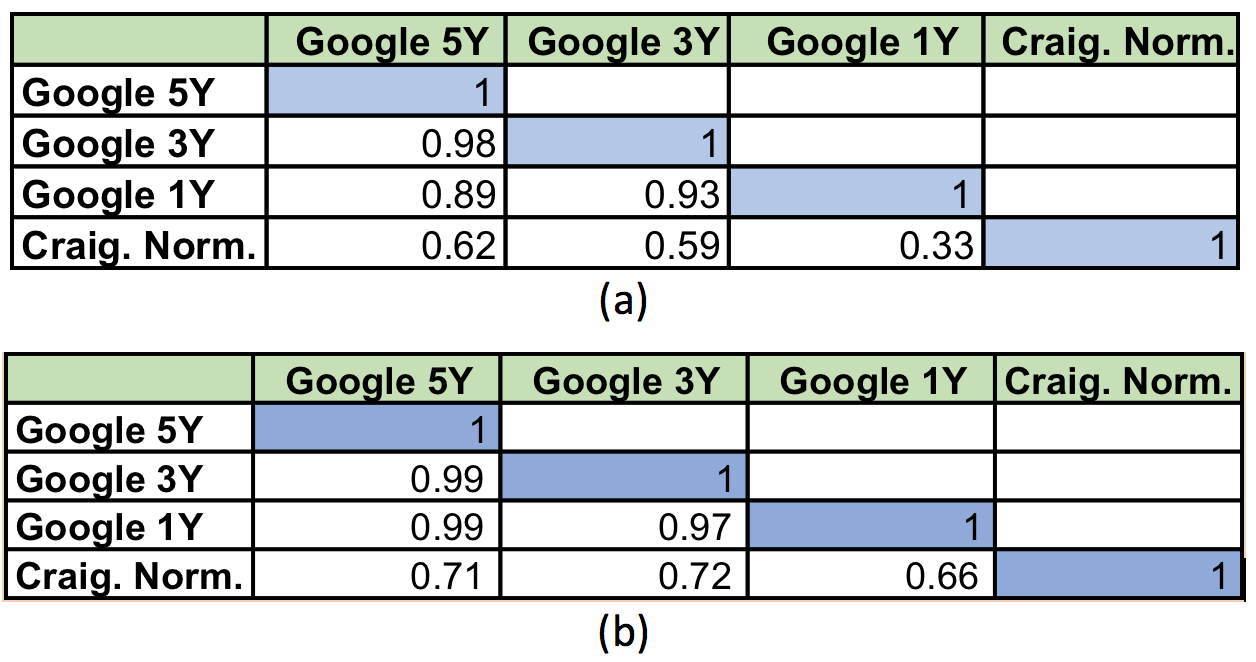}
\vspace{-0.1in}
\caption{\em Correlation table of Google Marijuana Keyword Search vs. Marijuana Related Corpora Ads in Craigslist (a) all states; (b) top 10 states}
\label{fig:google_craig}
\end{center}
\vspace{-0.2in}
\end{figure}

To validate and verify our data collected on Craigslist about marijuana use, we linked the ads to their geographic locations and used colored map to indicate the number of related keywords used across 50 states in the U.S. Fig. \ref{fig: states_with_4201} displays the marijuana state laws map as of May 2016 \cite{url2016governing} (Fig. \ref{fig: states_with_4201}(a)) and our  marijuana related corpora ads extracted from Craigslist (Fig. \ref{fig: states_with_4201}(b)). In the state marijuana laws map, light green, dark green and grey colors represent states with legalizing marijuana for recreation, for medical, and not legalizing, respectively. Fig. \ref{fig:google_craig}(a) and (b) show correlation tables of marijuana keyword search in Google over last five years (Google 5Y), three years (Google 3Y) and last year (Google 1Y) versus state population-normalized marijuana related ads in Craigslist of the 50 states and top 10 states. The data collected in Craigslist with ``marijuana'' related ads are strongly correlated with the states where marijuana uses are legal in some forms in both the state marijuana laws map and Google search. This is expected as marijuana uses are legal in those states; thus, lessors more often use marijuana related corpora (e.g., allowed or not allowed) in their ads to attract more lessees.  
\begin{figure}[t]
\begin{center}
\includegraphics[height = 1.5in, width = 3.5in]{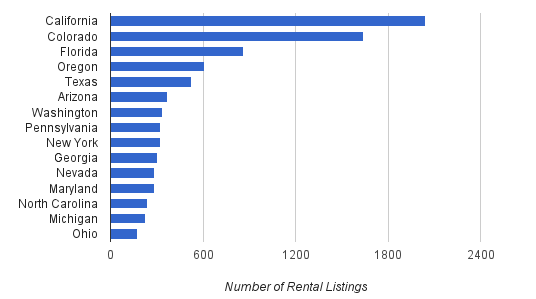}
\vspace{-0.2in}
\caption{\em  Frequencies of ``420 friendly" Corpora in the ``housing" Category}
\label{fig: states_with_420}
\end{center}
\vspace{-0.3in}
\end{figure}

\subsection{Marijuana Black Market: The Hidden World Revealed}

Interestingly, our data revealed that the terms ``420'' related words which have been used very often in the ads in some states such as Texas, Florida, or Nevada in Figure \ref{fig: states_with_420} where the use of marijuana in any form is illegal. Based on the federal laws and state marijuana laws map, we expected that less or no ads with concerns about the use of marijuana will be posted in those states because, obviously, no one in those areas is allowed to use marijuana in any forms. Surprisingly, our data indicated a contrary observation. Significant number of rental ads included marijuana related keywords in their ads have been posted in these areas. For example, Florida has more than 900 ads posted with marijuana related keywords. From the business and public health aspects, these ads suggest some level of actual marijuana uses in those communities; thus, the ads have been crafted with those keywords to attract more lessees in these areas. In other words, the evidence from the collected data can be interpreted as there exists some level of illegal marijuana uses in those states and the findings shed some light on those aspects that authorities and public health agencies should develop appropriate policy and regulation to restrain the reality of the marijuana use. 
\vspace{-0.1in}
\section{Conclusion}
\label{sec:conclusion} 

We have developed efficient and accurate text-mining algorithms to extract related marijuana corpora from unstructured ads in Craigslist to reveal the demographics of marijuana uses. Our data indicates strong correlation of ads with marijuana use concerns with the state marijuana laws map and Google marijuana keyword search in states where the marijuana use is legal. Interestingly, our data also reveals some strong concerns in marijuana uses in other states, where possession of marijuana in any form is illegal. The finding would suggest some hidden world of marijuana black market in those states. Our approach can be used as a marijuana surveillance tool for federal authorities and public health agencies in developing policy and regulations. 

\section{Acknowledgment}
The collected data is used solely for the research purpose. We neither relist nor resell the data to third parties.
\bibliographystyle{IEEEtran}
\bibliography{references}

\end{document}